\documentclass[twocolumn]{aastex61}

\submitjournal{ApJ}

\shorttitle{Energy dissipation of blazar}
\shortauthors{Fan, Wu \& Liao}

\begin{document}

\title{Constraints on the composition, magnetization, and radiative efficiency in jet of blazar}

\email{fanxl@hust.edu.cn, qwwu@hust.edu.cn, liaonh@pmo.ac.cn}

\author[0000-0002-0786-7307]{Xu-Liang Fan}
\affil{School of Physics, Huazhong University of Science and Technology, Wuhan 430074, China}
\affil{Guizhou Provincial Key Laboratory of Radio Astronomy and Data Processing, Guiyang 550025, China}

\author{Qingwen Wu}
\affiliation{School of Physics, Huazhong University of Science and Technology, Wuhan 430074, China}

\author{Neng-Hui Liao}
\affiliation{Key Laboratory of Dark Matter and Space Astronomy, Purple Mountain Observatory, Chinese Academy of Sciences, Nanjing 210008, China}



\begin{abstract}
  The composition and energy dissipation in jets are two of the fundamental questions of jet physics that are not fully understood. In this paper, we attempt to constrain the composition, magnetization as well as radiative efficiency for blazar with the recently released low-frequency radio catalog of the TIFR GMRT Sky Survey at 150 MHz. The jet power estimated from the low-frequency radio emission is much lower than that derived from spectral energy distribution fitting assuming one proton per electron. Assuming the jet power estimated from low-frequency radio emission is physical, the fraction of electron/positron pairs can be constrained with $n_{\rm pairs}/n_{\rm p} \sim 10$. By comparing the power carried by magnetic field and radiation with the jet power estimated from the low-frequency radio emission, we find both relatively high magnetization parameter of $\sigma \sim 0.5$ and radiative efficiency of $\eta \sim 0.4$ in the dissipation region of blazars. These results suggest that the magnetic reconnection processes may play an important role in the energy dissipation of blazars. We also explore the connection between these three parameters ($n_{\rm pairs}/n_{\rm p}$, $\sigma$, and $\eta$) and the black hole mass, disk luminosity as well as Eddington ratio. No significant correlation is found, except that $\sigma$ shows possible correlation with disk luminosity.
\end{abstract}




\section{Introduction}
Supermassive black holes coupled with accretion disks are currently widely accepted as central engines of active galactic nuclei (AGNs). A small fraction of AGNs are radio loud, which are believed to host relativistic jets~\citep{2017NatAs...1E.194P, 2017A&ARv..25....2P}. Until now, the jet formation and acceleration mechanisms are still fundamental open questions of jet physics~\citep{2007ApJ...658..815S, 2011ApJ...727...39M, 2013ApJ...764L..24S, 2016ApJ...833...30C}. Blandford-Znajek mechanism (BZ,~\citealt{1977MNRAS.179..433B}) and Blandford-Payne mechanism (BP,~\citealt{1982MNRAS.199..883B}) are two popular mechanisms for extracting energy of relativistic jets from central engines. For BZ mechanism, jet energy extraction is purely electromagnetic and directly coupled with the spin of black hole, which is supported by several observations in X-ray binaries (XRBs, e.g.,~\citealt{2012MNRAS.419L..69N}, but also see~\citealt{2013MNRAS.431..405R}). For BP mechanism, jet energy is extracted from the accretion disk through large-scale magnetic field anchored in accretion flow. Observationally, it is normally found that the radio emission appears in low-hard state of XRBs, while it will become weak or disappear in high-soft state~\citep{2001ApJ...554...43C, 2004MNRAS.355.1105F}, which suggests that the accretion mode also play a role in jet formation~\citep{2010ApJ...711...50T, 2013ApJ...765...62S, 2014ARA&A..52..529Y, 2016ApJ...817...71C}. In AGNs, the jet production efficiency, defined by the ratio between jet power and accretion power, is also found to be related to the black hole spin~\citep{2007ApJ...658..815S} and/or the geometrical thickness of the accretion disk~\citep{2016MNRAS.462..636A, 2017MNRAS.466.2294R}.

The energy in jet is mainly carried by three components, magnetic field, radiation and kinetic energy of matter. In principal, energy is dominated by Poynting flux at the base of jet~\citep{2015ApJ...803...30K}. Then magnetic energy converts into kinetic energy along the jet. Meanwhile the jet is accelerated~\citep{2010MNRAS.402..353L, 2011MmSAI..82...95K}. At a certain distance, the radiative particles are accelerated and most observed radiation (especially high energy emission) is related to this region. This region is also named dissipation region. There are some arguments that jet production efficiency can exceed unity under the magnetic arrested disk (MAD, e.g.,~\citealt{2011MNRAS.418L..79T, 2014Natur.515..376G}). The estimation of jet production efficiency is strongly dependent on the estimation of jet power. However, the deviation between the jet power estimated by different methods is large to one order of magnitude~\citep{2017MNRAS.465.3506P, 2018ApJS..235...39C}. Even for the nearest FR II radio galaxy Cygnus A, the accurate value of its jet power is under debate (see e.g.,~\citealt{2013ApJ...767...12G, 2016MNRAS.457.1124K} and reference therein).

The kinetic power, or the terminal jet velocity is strongly dependent on the composition of jet~\citep{2001Sci...291...84M}. But the jet content is difficult to constrain. Only the distribution of emitting particles can be constrained from observations, i.e., the non-thermal relativistic electrons~\citep{2006MNRAS.372.1741D}. But the number density of the thermal electrons and existence of protons can not be constrained from observations directly. There are several indirect methods to constrain the jet content for lobes of radio galaxies (e.g.,~\citealt{2006MNRAS.372.1741D, 2012ApJ...751..101K, 2016MNRAS.457.1124K}) and the dissipation regions of blazars in the literature (\citealt{1998Natur.395..457W, 2010MNRAS.409L..79G, 2014ApJS..215....5K, 2014ApJ...788..104Z}). Most of these methods are based on the constraint on the distribution of emitting electrons (especially the minimum energy of electrons $\gamma_{\rm min}$), and an independent method to estimate the kinetic jet power. Under the assumption energy equipartition or not, the power carried by other components (mainly protons) can be generally constrained through the comparison between the electron power and total kinetic power. Currently, most results of blazars showed that the matter was dominated by the electron/positron pairs on the number density, while the kinetic jet power was still dominated by protons~\citep{2012MNRAS.424L..26G, 2014ApJS..215....5K, 2016MNRAS.457.1124K, 2017MNRAS.465.3506P}. However, the observational and theoretical analyses of radio galaxies showed that jet power of FR II lobes could be dominated by leptons, while FR I lobes needed substantial protons to satisfy the pressure balance between the internal lobe and external environment (\citealt{2013ApJ...767...12G, 2018MNRAS.476.1614C}, also see~\citealt{2010MNRAS.402..497G} for similar discussions on blazars). In addition, there are some suggestions that the jet content is distinct for different jet powers~\citep{2006MNRAS.372.1741D}, or accretion modes~\citep{2013MNRAS.436..304P}. But it still lacks direct comparisons between jet content and central engines in the literature.

The magnetization parameter $\sigma$ is usually used to describe the level of the magnetic field dominance in jet with $\sigma = P_{\rm B}/P_{\rm m}$ (where $P_{\rm B}$ is the power carried by the magnetic field, $P_{\rm m}$ is the kinetic power of the material in jet). $\sigma$ is important on the acceleration mechanisms of the non-thermal electrons (internal shock or magnetic reconnection,~\citealt{2014ApJ...783L..21S, 2015SSRv..191..519S}). Moreover, there are several models to explain the short-term variability or hard $\gamma$-ray spectrum (e.g., mini-jet and current-driven instability,~\citealt{2009MNRAS.395L..29G, 2012MNRAS.427.2480N, 2014PhRvL.113o5005G}), which require the jet is strongly magnetized. The estimation of $\sigma$ in blazar is usually based on spectral energy distribution (SED) fitting.~\citet{2010MNRAS.402..497G} and~\citet{2013ApJ...774L...5Z} calculated the power carried by each component (namely radiation, magnetic field, non-thermal electrons, and protons whose number density is assumed to be equal to electrons) in jet based on the results of SED fitting.~\citet{2013ApJ...774L...5Z} found that $\sigma$ of flat spectrum radio quasar (FSRQ) was close to unity, while the fraction of jet power carried by magnetic field was much smaller for luminous blazars (mainly FSRQs) in~\citet[their Figure 5]{2010MNRAS.402..497G}. The deviation may be mainly due to the different minimum energies of emitting electrons applied in these two works~\citep{2013ApJ...774L...5Z}. Based on the SED fitting, it was also found that $\sigma$ was different for different subclasses of blazars, i.e., FSRQs and BL Lac objects (BL Lacs). But some authors showed that $\sigma$ was larger for FSRQs than that of BL Lacs~\citep{2013ApJ...774L...5Z, 2018ApJS..235...39C}, while other results were opposite~\citep{2010MNRAS.402..497G}. Therefore, estimation of $\sigma$ independent on the distribution of emitting electrons is needed.

The radiative efficiency $\eta$ describes the fraction of the jet power dissipated into radiation with $\eta = P_{\rm rad}/P_{\rm j}$ (where $P_{\rm rad}$ is the power carried by the radiation, $P_{\rm j} = P_{\rm m}+P_{\rm rad}+P_{\rm B}$ is total power in jet). The radiative efficiency of some $\gamma$-ray bursts (GRBs) are found to be as high as 90\%~\citep{2007ApJ...655..989Z}, which is much larger than the prediction of the standard shock model (\citealt{2011ApJ...726...90Z} and references therein). These high efficiencies suggest the energy dissipation by magnetic reconnection in jet may be important~\citep{2011ApJ...726...90Z}. The radiative efficiency of blazar was also found to be larger than 10\%~\citep{2012Sci...338.1445N, 2013ApJ...774L...5Z}, which is also difficult to produce under the internal shock model~\citep{2011ApJ...726...90Z}.

In this paper, we constrain the jet content, magnetization parameter and radiative efficiency with the jet power estimations independent on the distribution of emitting electrons. We also compare the trends of all these three parameters with black hole mass, disk luminosity and Eddington ratio. In section 2, we present the method and data used in our work. Section 3 shows the results of the distribution and correlation analyses. We discuss the implications of our results in section 4. In section 5, we summarize the main results. In this paper, we use a $\Lambda$CDM cosmology with $H_0 = 70$ km s$^{-1}$ Mpc$^{-1}$, $\Omega_{\rm M}$=0.3, $\Omega_{\Lambda}$=0.7, consistent with~\citet{2014Natur.515..376G} and ~\citet{2015MNRAS.448.1060G}.

\section{Method and Data}
Using the multi-wavelength SED fitting of blazar, one can derive the distribution of emitting electrons and the magnetic field strength in dissipation region. With the assumption one proton per emitting electron in jet, the power carried by radiation, relativistic electrons, magnetic field and protons can be estimated for blazar zone (see~\citealt{2008MNRAS.385..283C} and~\citealt{2014Natur.515..376G} for more details). The total jet power is the sum of these components with $P_{\rm fit} = P_{\rm B}+P_{\rm p}+P_{\rm e}$, where $P_{\rm B}$, $P_{\rm p}$ and $P_{\rm e}$ is the power carried by the magnetic field, proton and electron, respectively.

Direct measurements of the kinetic energy in jet is derived from the observations of the large-scale structures of radio galaxies~\citep{2006MNRAS.372.1741D, 2008ApJ...686..859B, 2012MNRAS.423.2498D, 2012ApJ...751..101K}. Based on the direct measurements of jet power, some empirical relations between kinetic power and extended radio emission are built~\citep{1999MNRAS.309.1017W, 2005ApJ...623L...9P, 2007MNRAS.381..589M, 2010ApJ...720.1066C, 2011ApJ...735...50W, 2013ApJ...767...12G, 2017MNRAS.467.1586I}. Although there are some arguments that the $P_{\rm kin}$-$L_{151}$ relation has dependence on the cluster environment~\citep{2013MNRAS.430..174H}, age of radio lobes~\citep{2018MNRAS.475.2768H}, or the composition of jet~\citep{2013ApJ...767...12G}, some theoretical and observational attempts found that there was an uniform relation for all active radio lobes~\citep{2013ApJ...767...12G, 2018MNRAS.475.2768H}.

Given the jet power estimated from low-frequency radio emission is correct, and the energy losses between the dissipation region and the large scale (extended) jets only take place via blazar radiation (ignore the radiation losses of large scale jets), the discrepancy between the estimated jet power from SED fitting and low-frequency radio emission means the assumption one proton per electron overestimates the jet power from SED fitting~\citep{2014ApJS..215....5K, 2016Galax...4...12S, 2017MNRAS.465.3506P}. Thus the ratio between electron/positron pairs and protons can be constrained with~\citep{2016Galax...4...12S}
\begin{equation}
\begin{centering}
\frac{n_{\rm pairs}}{n_{\rm p}} = \frac{1}{2}(\frac{P_{\rm p}}{P_{\rm kin} - P_{\rm B} - P_{\rm e}}-1).
\label{com}
\end{centering}
\end{equation}
The magnetization parameter $\sigma$ is estimated with $P_{\rm B}$ from SED fitting and the jet power from low-frequency radio emission with $\sigma = P_{\rm B}/(P_{\rm kin} - P_{\rm B})$. Radiative efficiency $\eta$ of blazar can be calculated with $P_{\rm rad}/(P_{\rm kin}+P_{\rm rad})$.

Recently, several new low-frequency radio surveys released their source catalogs, such as the TIFR GMRT Sky Survey (TGSS) with Giant Metrewave Radio Telescope (GMRT,~\citealt{2017A&A...598A..78I}), GaLactic and Extragalactic All-sky MWA (GLEAM) survey with Murchison Widefield Array (MWA,~\citealt{2017MNRAS.464.1146H}), and LOFAR Two-metre Sky Survey (LoTSS) with International Low-Frequency Array (LOFAR,~\citealt{2017A&A...598A.104S}). These recently released catalogs from low-frequency radio surveys give us good opportunities to explore the jet power for large sample of blazars.

~\citet{2014Natur.515..376G} fitted the SEDs of 217 blazars based on one-zone leptonic model and calculated the power carried by each component ($P_{\rm rad}$, $P_{\rm B}$, $P_{\rm p}$, and $P_{\rm e}$). They also listed three groups of black holes mass which were based on the virialized estimation of three emission lines (H$\beta$, Mg{\sc ii} and C{\sc iv}), respectively. In our work, the black hole mass estimations based on the H$\beta$ measurements are favored, as H$\beta$ is the best calibrated line with reverberation mapping method. The data based on C{\sc iv} are least favored, as the calibration of C{\sc iv} is less reliable~\citep{2011ApJS..194...45S}. The disk luminosity estimated by line luminosity in~\citet{2014Natur.515..376G} is used. The Eddington ratio is calculated with $L_{\rm disk}/L_{\rm Edd}$, where $L_{\rm Edd}$ is Eddington luminosity. We cross-match the TGSS ADR1 catalog~\citep{2017A&A...598A..78I} with the sample of~\citet{2014Natur.515..376G} within the distance of 3\arcsec. This results in a sample of 133 objects. We estimate their jet power with the 150 MHz radio flux from TGSS ADR1 and the $P_{\rm kin}$-$L_{151}$ relation in~\citet{2013ApJ...767...12G}
\begin{equation}
\begin{centering}
P_{\rm kin} = 3 \times 10^{44} (\frac{L_{151}}{\rm 10^{25}~W~Hz^{-1}~sr^{-1}})^{0.67} ~~\rm erg~s^{-1}.
\label{pext}
\end{centering}
\end{equation}

\section{Results}
\begin{figure}
\begin{center}
\includegraphics[angle=0,scale=0.38]{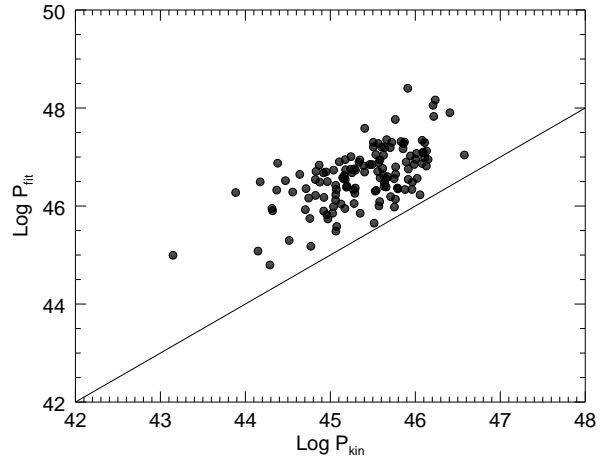}
\caption{The jet power estimation from low-frequency radio emission and SED fitting. The solid line shows the equation line. The jet power estimated from SED fitting is larger than that from low-frequency radio emission for all the 133 objects. \label{pj}}
\end{center}
\end{figure}

\begin{figure*}
\begin{center}
\includegraphics[angle=0,scale=0.29]{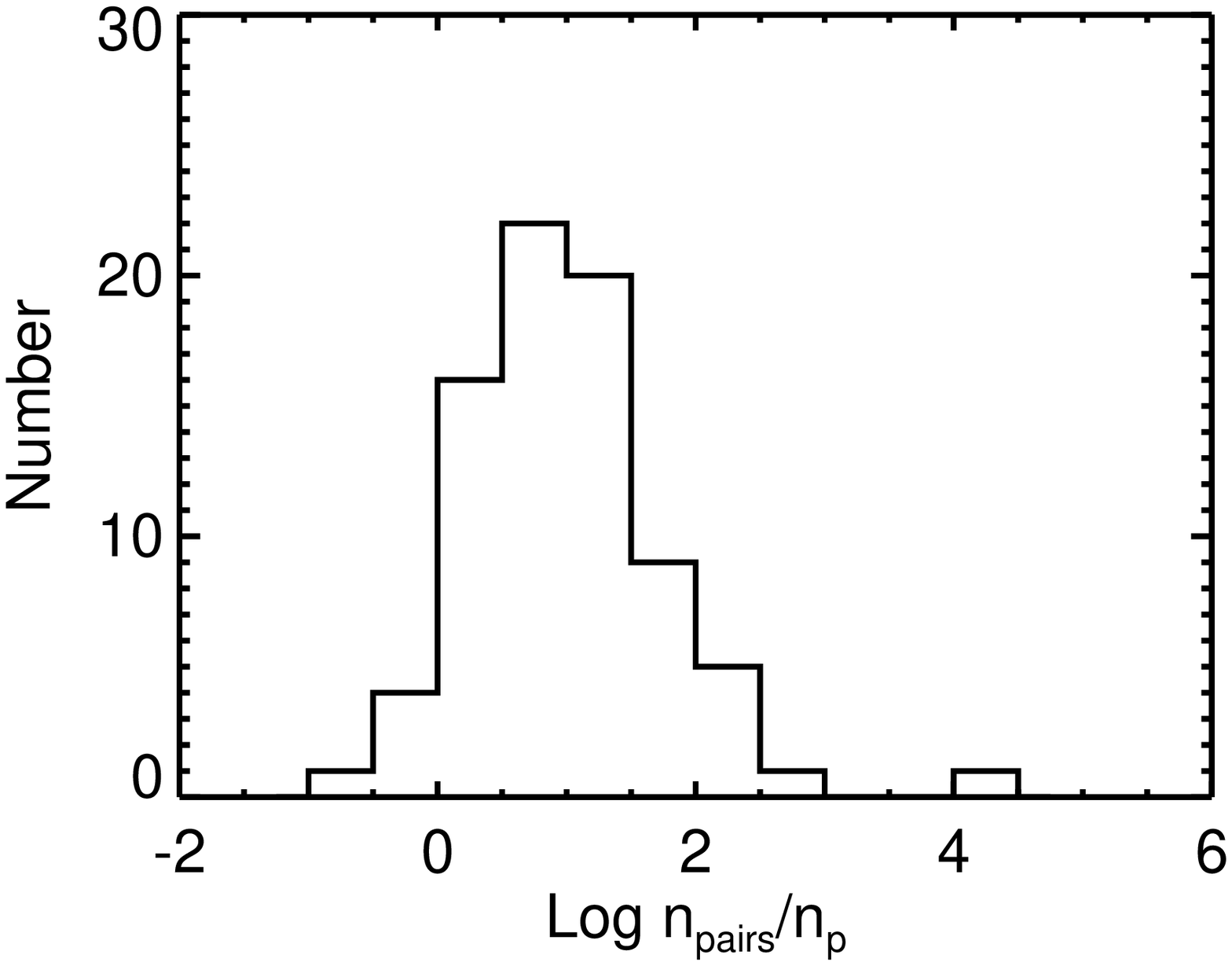}
\includegraphics[angle=0,scale=0.29]{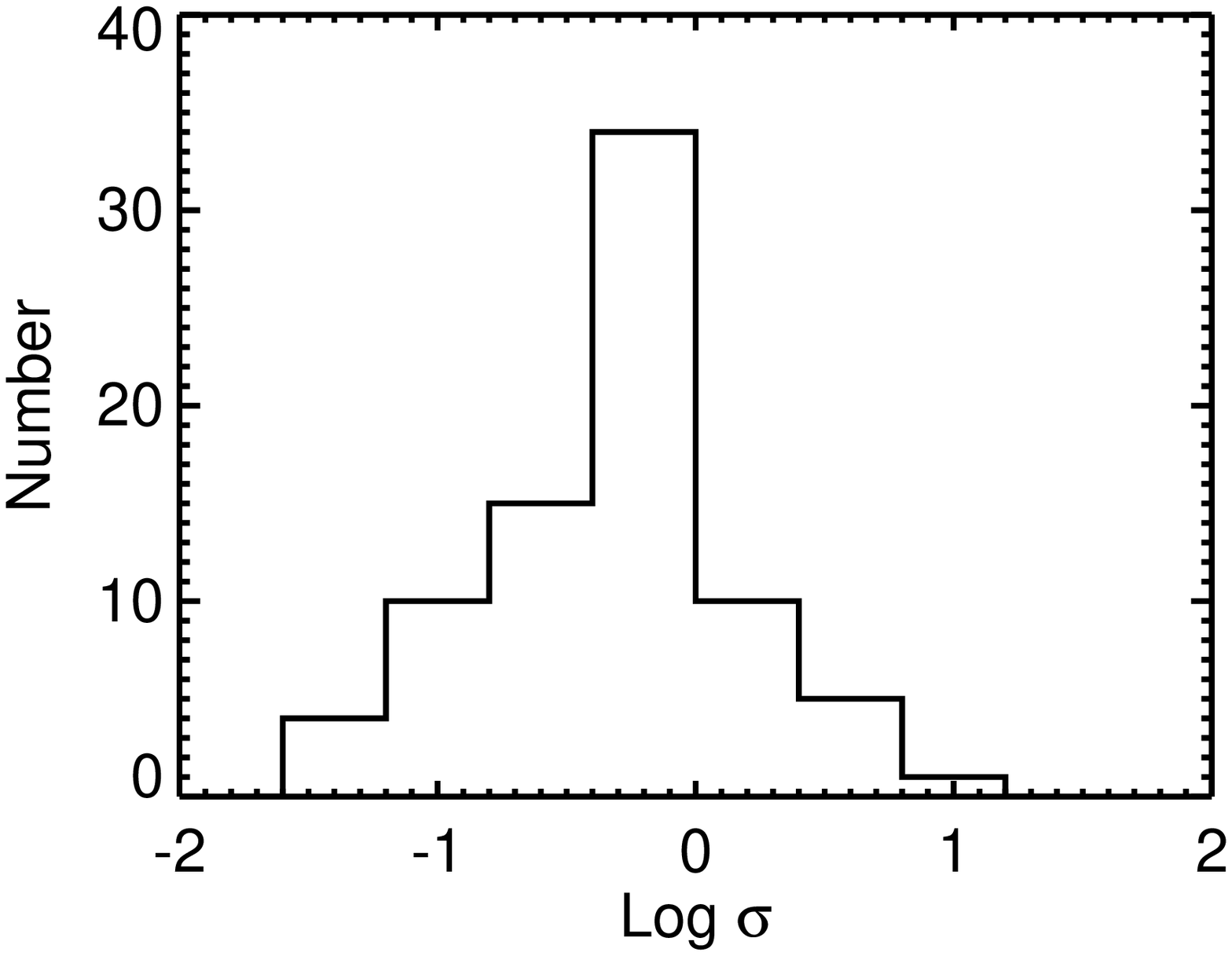}
\includegraphics[angle=0,scale=0.29]{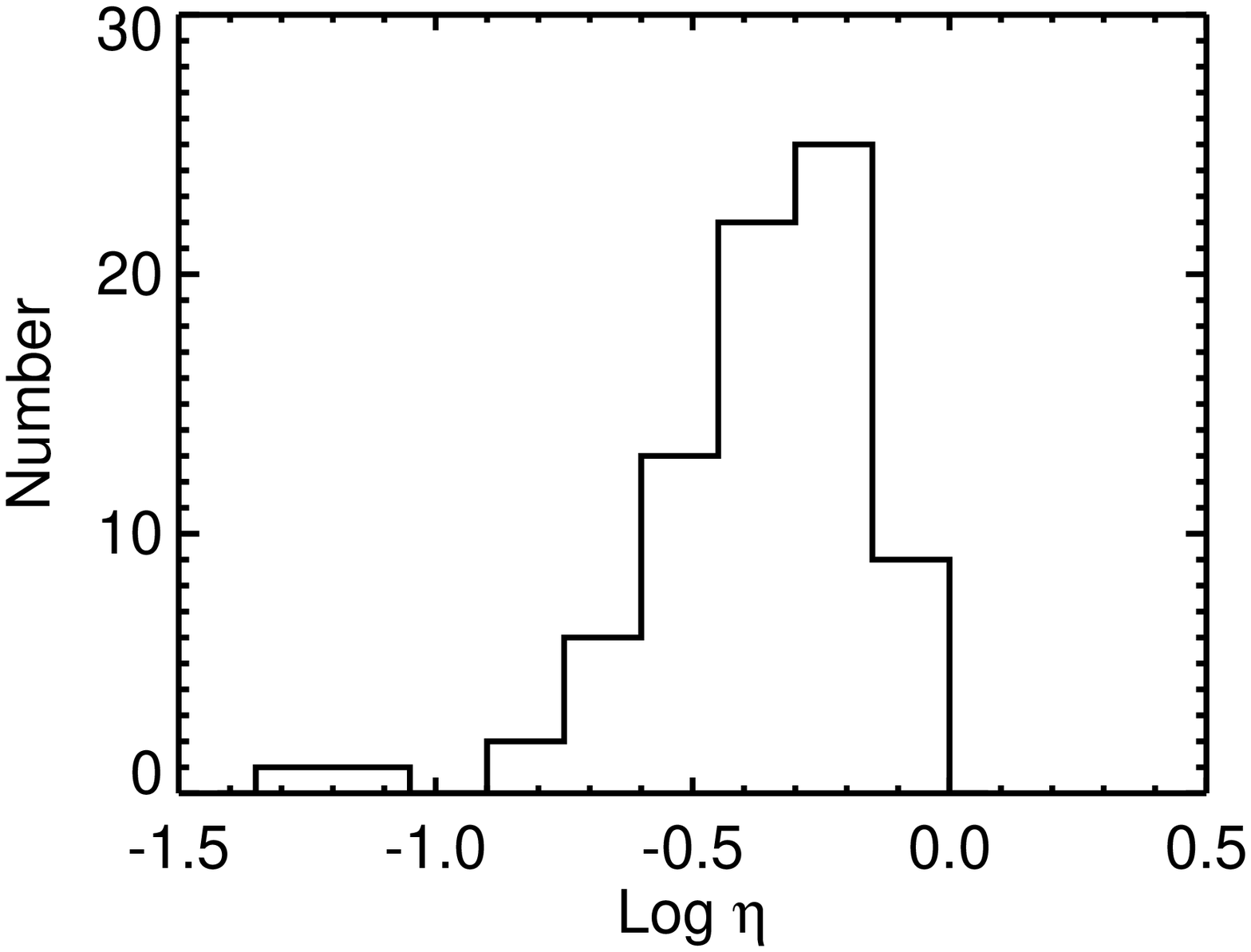}
\caption{From left to right are the distributions of $n_{\rm pairs}/n_{\rm p}$, $\sigma$, and $\eta$, respectively. \label{dis}}
\end{center}
\end{figure*}
The jet power estimated from SED fitting versus that from low-frequency radio emission is plotted in Figure~\ref{pj}. It shows that $P_{\rm fit} > P_{\rm kin}$ for all the 133 blazars, which means that the assumption one proton per electron indeed overestimates the jet power in the emission zone of blazar. Among these 133 objects, there are 79 objects which we can calculate the fraction of electron/positron pairs with Equation~\ref{com}. The unreasonable results for remaining objects can be due to the uncertainties of jet power estimation either from the SED fitting or the $P_{\rm kin}$-$L_{151}$ relation, or the energy losses from pc to kpc scale during the growth of extended structures (we will discuss this in details in Section 4). In the following analyses, we just consider the 79 objects with composition estimations. The detailed information of these 79 objects are listed in Table~\ref{t2} in the Appendix.

The distributions of $n_{\rm pairs}/n_{\rm p}$, $\sigma$ and $\eta$ for our sample are shown in Figure~\ref{dis} (also see Table~\ref{t2} in the Appendix). The mean value of $n_{\rm pairs}/n_{\rm p}$ is 9.79, with majority concentrates in range 1 to 100 (left panel of Figure~\ref{dis}). $\sigma$ spans in the range 0.03 to 7.86, with the mean value 0.49 (middle panel of Figure~\ref{dis}). The mean value of radiative efficiency $\eta$ is 0.42 with the range 0.05 to 0.94 (right panel of Figure~\ref{dis}).

We further explore the evolution of jet content, magnetization parameter, and radiative efficiency along with the black hole mass, disk luminosity and Eddington ratio. Before correlation analysis, we firstly compare the distributions of black hole mass, disk luminosity and Eddington ratio between our sample and the sample of~\citet{2014Natur.515..376G}. The Kolmogorov-Smirnov (K-S) tests show no evidences for distinct distributions between our sample and the Ghisellini's (with probabilities larger than 0.22), which suggest that our results can reflect the general trends for the parent sample in~\citet{2014Natur.515..376G}.
\begin{table}
\begin{center}
  \caption{The results of correlation analysis. Columns 1 and 2 are the two parameters which are applied for correlation analysis, respectively. $\rho$ is the correlation coefficient of the Spearman correlation test. P is the chance probability of no correlation. The last column lists the percentage from the bootstrapping technique for which can be considered as correlation.}
  \setlength{\tabcolsep}{3pt}
  \label{cor}
  \begin{tabular}{cccccc}
  \hline
  Par A & Par B & $\rho$ & P & Per\\
  \hline
  $n_{\rm pairs}/n_{\rm p}$ & $M_{\rm BH}$ & 0.09 & 0.41 & 0.01\\
  $n_{\rm pairs}/n_{\rm p}$ & $L_{\rm disk}$ & 0.07 & 0.56 & 0.01\\
  $n_{\rm pairs}/n_{\rm p}$ & $L_{\rm disk}/L_{\rm Edd}$ & -0.08 & 0.51 & 0.01\\
  $\sigma$ & $M_{\rm BH}$ & 0.20 & 0.08 & 0.16\\
  $\sigma$ & $L_{\rm disk}$ & 0.22 & 0.05 & 0.24\\
  $\sigma$ & $L_{\rm disk}/L_{\rm Edd}$ & -0.07 & 0.57 & 0.01\\
  $\eta$ & $M_{\rm BH}$ & 0.14 & 0.21 & 0.06\\
  $\eta$ & $L_{\rm disk}$ & 0.13 & 0.24 & 0.05\\
  $\eta$ & $L_{\rm disk}/L_{\rm Edd}$ & -0.06 & 0.57 & 0.01\\
  \hline
  \end{tabular}
\end{center}
\end{table}

The Spearman rank correlation test is used to explore the correlations. It is taken as correlation when the significance is larger than 95\% (P $< 0.05$). We also perform a bootstrapping technique to evaluate the reliability of the correlation results. We randomly draw ($1-1/e$) objects from our sample and derive the correlation results for the sub-sample, this process is performed for 10000 times. We record the percentages which could be considered as correlations~\citep{1998PASP..110..660P}.

\begin{figure*}
\begin{center}
\includegraphics[angle=0,scale=0.39]{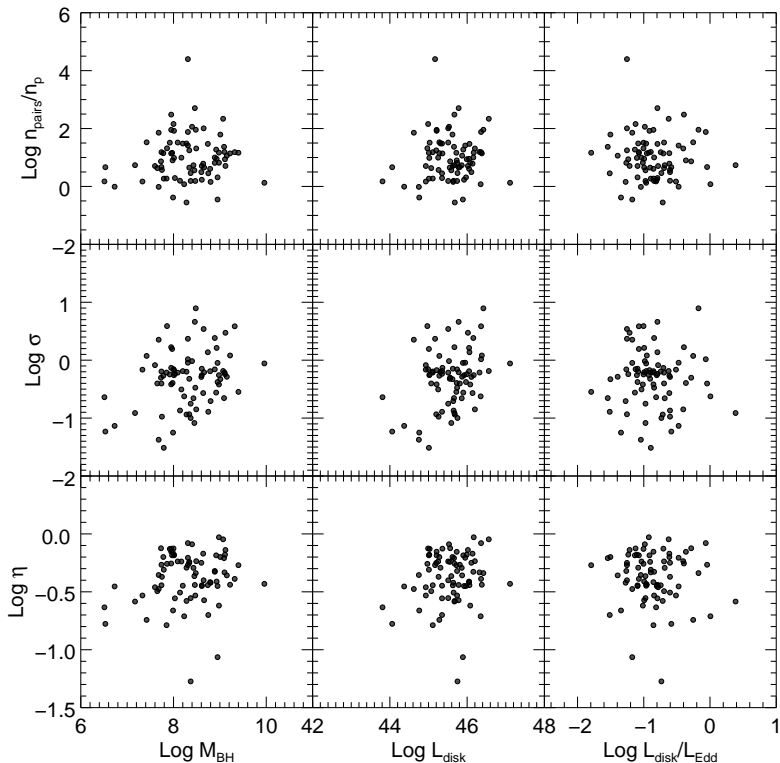}
\caption{The correlation between energy dissipations of jet and central engines. Left: Black hole mass. Middle: disk luminosity. Right: Eddington ratio. The top label is the correlation between $n_{\rm pairs}/n_{\rm p}$ and central engines. The middle label is the correlation between $\sigma$ and central engines. The bottom label is the correlation between $\eta$ and central engines. \label{pic}}
\end{center}
\end{figure*}

The top panels of Figure~\ref{pic} show the scatters of $n_{\rm pairs}/n_{\rm p}$ and central engines. The results of the correlation test are summarized in Table~\ref{cor}. No correlation is shown between $n_{\rm pairs}/n_{\rm p}$ and black hole mass, disk luminosity, or Eddington ratio. Table~\ref{cor} also lists the percentages from the bootstrapping technique. The results of bootstrapping confirm that no correlation exists between composition and central engine.

It seems that $\sigma$ has weak correlation with disk luminosity (middle panels of Figure~\ref{pic}, Table~\ref{cor}). The results of bootstrapping find that 24\% of the sub-samples can derive correlations between these two parameters. No correlation is found for $\sigma$ versus black hole mass, and $\sigma$ versus Eddington ratio. The radiative efficiency also shows no correlation with black hole mass, disk luminosity, and Eddington ratio (bottom panels of Figure~\ref{pic}, Table~\ref{cor}).

\section{Discussions}
\citet{2013ApJ...767...12G} defined a normalization factor $g$ to reflect the uncertain physics in lobes, such as the composition, magnetic field strength, electron spectrum, jet speed, electron Lorentz factor. The value of $g$ can vary from 1 to 8 for different situations (their Figure 1). As suggested by~\citet{2013ApJ...767...12G} from some observational concerns, $g = 2$ is used in our Equation~\ref{pext}.~\cite{1999MNRAS.309.1017W} also defined $f$ factor to consider the similar uncertainties in $P_{\rm kin}$-$L_{151}$ relation. They argued that $1 \leq f \leq 20$. For our analysis here, another uncertainty comes from the energy losses which could be used to form the large scale structure after the power leaving the dissipation region of blazar. This can lead to an underestimation of $P_{\rm kin}$ in blazar region about a factor of a few~\citep{2013MNRAS.430..174H}. Considering these uncertainties for different conditions, we try several other relations in the literature for our analyses. Firstly, we add another factor of 2 in Equation~\ref{pext} to account for the underestimation due to energy losses from structure expanding. The number of sources whose pair fraction can be calculated increases to 110~\footnote{The uncertainty of jet power for a small parts of the 133 objects may come from the SED fitting, e.g., the electron spectrum~\citep{2014ApJ...788..179C}, or the duty cycle of blazar activity ~\citep{2017MNRAS.467.4565L}.}. The mean values of $n_{\rm pairs}/n_{\rm p}$, $\sigma$ and $\eta$ are 5.16, 0.33 and 0.33, respectively. Interestingly, we note that the correlation between $\sigma$ and disk luminosity gets more significant, with P = 1.50$\times 10^{-4}$. Then we consider the relation $P_{\rm kin} = f^{3/2} 3 \times 10^{45} (\frac{L_{151}}{\rm 10^{28}~W~Hz^{-1}~sr^{-1}})^{6/7}$ in~\citet{1999MNRAS.309.1017W} with $f = 5$~\citep{2018MNRAS.475.2768H} and $f = 10$, and $P_{\rm kin} = 5 \times 10^{46} (\frac{L_{151}}{\rm 10^{28}~W~Hz^{-1}~sr^{-1}})^{0.89}$ in~\citet{2017MNRAS.467.1586I}. The mean values of $n_{\rm pairs}/n_{\rm p}$, $\sigma$ and $\eta$ range from 6.71 to 12.59, 0.39 to 0.59, and 0.35 to 0.47, respectively. The results are well consistent with those from Equation 2.

Figure~\ref{pj} shows that the jet power estimated from the low-frequency radio emission is lower than that from SED fitting assuming one proton per electron. Assuming the jet power estimated from the low-frequency radio emission is correctly reflect the real jet power, we constrain the jet composition by comparing the jet power derived by these two methods. The results manifest that the jet of blazar contains an important fraction of eletron/positron pairs, with $n_{\rm pairs}/n_{\rm p} \sim 10$ (left panel in Figure~\ref{dis}). Our results are consistent with previous works, where they pointed out that the electron/positron pairs were important on the number density in blazar jet~\citep{2014ApJS..215....5K, 2016Galax...4...12S, 2017MNRAS.465.3506P}. The ratio 10 pairs per proton is also consistent with the upper limit for pairs to avoid the Compton rocket effect~\citep{2010MNRAS.409L..79G, 2012MNRAS.424L..26G}. In addition, no evidence is found for evolution of the pairs along black hole mass, disk luminosity, or Eddington ratio (Figure~\ref{pic}, Table~\ref{cor}).

We find that the contents of jets are independent on Eddington ratio, although most objects in our sample are FSRQs accreted under radiative efficient state.~\citet{2018MNRAS.476.1614C} also obtained similar conclusion for FR II galaxies.~\citet{2016RAA....16..173F} found that the intrinsic jet power $P_{\rm kin}/\Gamma^2$ (where $\Gamma$ is the bulk Lorentz factor of jet) for FSRQs and BL Lacs had similar distributions, which also supported that the material energy of jets was independent on accretion mode.

Using the jet power estimated from low-frequency radio emission and the magnetic power estimated from the SED fitting, we find $\sigma \sim 0.5$ in the dissipation region of blazars (middle panel in Figure~\ref{dis}). In strongly magnetized shock ($\sigma \gtrsim 10^{-3}$), the particle acceleration would be suppressed (\citealt{2015SSRv..191..519S}, but also see~\citealt{2017MNRAS.464.4875B}). Meanwhile, the acceleration through the magnetic reconnection would be efficient~\citep{2014ApJ...783L..21S}. Our results indicate that the magnetic reconnection may be the main source to accelerate non-thermal particles in blazars~\citep{2016Galax...4...12S}.~\citet{2015MNRAS.449..431J} computed the theoretical SEDs of blazars and concluded that $\sigma \ll$ 1 to satisfy the observed Compton dominance for FSRQs (also see~\citealt{2014ApJ...796L...5N}). Their results were based on $n_p/n_e = 1$. Once eletron/positron pairs exist in jets, the jet production efficiency would reduce. Therefore the high Compton dominance can be produced even $\sigma \gtrsim 1$~\citep{2016Galax...4...12S}.

The magnetization parameter $\sigma$ shows a possible correlation, although very weak with disk luminosity (Figure~\ref{pic}, Table~\ref{cor}). If this is intrinsic, it would suggest that the dissipation region in brighter disk is closer to the central engine, because $\sigma$ decreases with the distance from the central object with $\sigma \propto r^{-\alpha}$ (although the detailed value of $\alpha$ is under debate, e.g.,~\citealt{2011ApJ...729..104K, 2011MNRAS.411.1323G, 2017ApJ...836..241T}). This can be caused by the increasing cooling with the higher disk radiation due to the external Compoton process.

The radiative efficiency $\eta \sim 0.4$ for blazars, which is much larger than the prediction of internal shock model (generally less than 15\%). For magnetic reconnection process, $\eta$ can be as high as 90\%~\citep{2011ApJ...726...90Z}. The high efficiency also suggests that magnetic reconnection is important to power the radiation of blazars, which is consistent with the strong magnetization in jet.

It also needs to note other possibilities for the discrepancy of the jet power estimations between different methods. These include that the jet power estimated by the low-frequency radio emission is underestimated due to the intermittent activity of jet~\citep{2016Galax...4...12S}, or the remnant sources in the sample~\citep{2018MNRAS.475.2768H}. If this is the fact, $n_{\rm pairs}/n_{\rm p}$, $\sigma$ and $\eta$ should be overestimated. But for our sample, the Eddington ratio is still high ($> 0.01$). And most sources are FSRQs which are unified with long-lived FR IIs hosted enhanced large-scale radio structures~\citep{1995PASP..107..803U}. No signature is shown for the transition on the accretion modes or dying extended structures.

\section{Summary}
The composition, magnetization and radiative efficiency are important to constrain the mechanisms of jet formation, particle acceleration and energy dissipation of blazar. In this work, we explore these issues with the recently released TGSS ADR1 catalog at 150 MHz. Our results manifest that, 1) The leptons are dominated on number density in blazar zone, with the ratio between electron/positron pairs and protons about 10. 2) The magnetization parameter of blazar is close to unity. The radiative efficiency is about 40\%, which is much larger than prediction of shock model. Both the strong magnetization and high efficiency suggest that magnetic reconnection process may be important to power the radiation of blazars. 3) No significant correlation is found between the composition, magnetization parameter, radiative efficiency and black hole mass, disk luminosity as well as Eddington ratio, except that the magnetization parameter shows possible correlation with disk luminosity.

\acknowledgments
We are grateful to the anonymous referee for constructive comments and suggestions that greatly improved this manuscript. We thank Xin-Wu Cao and Liang Chen for useful discussions. This research has made use of the VizieR catalogue access tool, CDS, Strasbourg, France. The original description of the VizieR service was published in A\&AS 143, 23. This research is supported by National Natural Science Foundation of China (NSFC; grants 11573009, 11622324 and 11703093) and Guizhou Provincial Key Laboratory of Radio Astronomy and Data Processing.

\vspace{5mm}
\facilities{GRMT (TGSS ADR1)}

%


\bibliographystyle{aasjournal}
\bibliography{bib}

\appendix
\begin{longrotatetable}
\begin{deluxetable*}{cccccccccccccccccc}
\tablecaption{The source information. \label{t2}}
\tabletypesize{\scriptsize}
\setlength{\tabcolsep}{2pt}
\tablehead{
\colhead{Name} & \colhead{RA} &
\colhead{DEC} & \colhead{z} & \colhead{M$_{\rm BH}$} &
\colhead{L$_{\rm disk}$} & \colhead{L$_{\rm disk}$/L$_{\rm Edd}$} &
\colhead{F$_{150}$} & \colhead{L$_{150}$} &
\colhead{P$_{\rm kin}$} &
\colhead{P$_{\rm B}$} & \colhead{P$_{\rm e}$} &
\colhead{P$_{\rm P}$} & \colhead{P$_{\rm fit}$} &
\colhead{P$_{\rm rad}$} & \colhead{$n_{\rm pairs}/n_{\rm p}$} &
\colhead{$\sigma$} &
\colhead{$\eta$} \\
 & & & & (M$_{\odot}$) & (erg s$^{-1}$) & & (mJy) & (erg s$^{-1}$ Hz$^{-1}$ sr$^{-1}$) &
 (erg s$^{-1}$) & (erg s$^{-1}$) & (erg s$^{-1}$) & (erg s$^{-1}$) &
 (erg s$^{-1}$) & (erg s$^{-1}$) & & &
}
\startdata
0004-4736  &  0.07655 & -47.60567 &  0.880 &  7.85 & 45.11 &  -0.85  &    664.90 & 33.30 & 45.35 & 44.93 & 43.98 & 45.79 & 45.85 & 44.64  &  0.27 &  -0.21 &  -0.79 \\
0023+4456  &  0.39320 &  44.94397 &  2.023 &  7.78 & 45.28 &  -0.61  &    211.80 & 33.70 & 45.62 & 45.16 & 44.88 & 46.75 & 46.77 & 45.85  &  1.15 &  -0.27 &  -0.20 \\
0042+2320  &  0.70125 &  23.33396 &  1.426 &  9.01 & 45.62 &  -1.50  &    342.80 & 33.54 & 45.51 & 45.01 & 44.96 & 47.20 & 47.21 & 45.74  &  1.79 &  -0.33 &  -0.20 \\
\enddata
\tablecomments{Column 1 is the source name. Column 2-3 give the coordinates of each object. The redshift is given in column 4. Column 5-7 are the black hole mass, disk luminosity and Eddington ratio taken from ~\citet{2014Natur.515..376G}. Column 8-9 are the 150 MHz radio flux and radio luminosity derived from TGSS survey. Column 10 is the jet power estimated by equation~\ref{pext} in this work. Column 11-13 are the jet power carried by magnetic field, electrons, and protons, respectively, which are derived from ~\citet{2014Natur.515..376G}. Column 14 is the total jet power derived from SED fitting. Column 15 is the jet power carried by radiation. Column 16 is the fraction of elctron/positron pairs. Column 17 is the magnetization parameter. Column 18 is the radiative efficiency. All the values, except redshift and 150 MHz radio flux, are in logarithmic space. This table is available in its entirety in machine-readable form.}
\end{deluxetable*}
\end{longrotatetable}

\end{document}